\documentclass[conference]{IEEEtran}
\IEEEoverridecommandlockouts
\usepackage{cite}
\usepackage{amsmath,amssymb,amsfonts}
\usepackage{algorithmic}
\usepackage{graphicx}
\usepackage{textcomp}
\usepackage{xcolor}
\usepackage{optidef}
\usepackage{array}

\newcolumntype{C}[1]{>{\centering\let\newline\\\arraybackslash\hspace{0pt}}m{#1}}

\allowdisplaybreaks
\def\BibTeX{{\rm B\kern-.05em{\sc i\kern-.025em b}\kern-.08em
    T\kern-.1667em\lower.7ex\hbox{E}\kern-.125emX}}

\begin{document}
\title{Optimal Design and Cascading Failure Evaluation \\ of Remedial Action Schemes\\
}

\author{\IEEEauthorblockN{Aditya Rangarajan and Line Roald}
\IEEEauthorblockA{\textit{Dept. of Electrical and Computer Engineering, University of Wisconsin-Madison}, Madison, Wisconsin, USA \\
\{arangarajan4, roald\}@wisc.edu
}
% \and
}

\maketitle

\begin{abstract}
% This document is a model and instructions for \LaTeX.
% This and the IEEEtran.cls file define the components of your paper [title, text, heads, etc.]. *CRITICAL: Do Not Use Symbols, Special Characters, Footnotes, 
% or Math in Paper Title or Abstract.
Remedial action schemes (RAS) are often seen as an alternative to building new transmission infrastructure to relieve congestion in the system. Consequently, there has been a rapid growth in the number of RAS in electric power systems across the world. However, most RAS rely on fixed parameters and hence cannot adapt to the rapidly evolving nature of the electric grid. In this paper, an optimization framework (RAS-SCOPF) to automate the RAS design procedure is proposed. The proposed framework is a mixed integer quadratic program (MIQP) that chooses a set of optimal RAS actions and minimizes load shed when a contingency occurs. The cost of operation of the RAS-SCOPF is compared against those of standard OPF and SCOPF formulations. Moreover, the risk of cascading failure for the different formulations are evaluated using a DC power flow based cascading failure simulator (CFS). The proposed framework is applied to the RTS-96 24-bus network. The inclusion of RAS allows the system to be operated at a lower cost while preventing any contingency from evolving into cascading blackouts. 
\end{abstract}


\section{Introduction}
The introduction of competitive electricity markets along with increasing electricity demand and renewable generation have led to increased stress on the existing transmission infrastructure. As a result, the electric power system is forced to operate closer to its limits. Since post-contingency security constraints are often the source of congestion, 
there is an increasing reliance on post-contingency control avoid post-contingency overloads and maintain secure system operation \cite{opf}. 
% While this is an economically beneficial way to increase power transfer capacity, it also makes system operations more complex and prone to errors.
Post-contingency manual corrective actions of the operator (e.g., generator redispatch, adjusting transformer tap settings, etc) can be too slow to arrest the propagation of disturbances. This has led to the global adoption of fast-acting system-wide protection systems, called Remedial Action Schemes(RAS), to maintain reliability \cite{survey}. 

According to the North American Electric Reliability Corporation (NERC), RAS are automatic protection systems that detect abnormal system conditions and take predetermined and fast control actions, including  but not limited to, generator rejection, load shedding, and line switching \cite{nerc}. RAS are also commonly referred to as system integrity protection schemes (SIPS) or special protection systems (SPS). RAS can use measurements from and take remedial actions at remote locations of the system, and thus differ from local protection systems. Since RAS can reduce violations of post-contingency constraints, %RAS can allow operators to relax N-1 security constraints and thus allow for an increase in power transfer capacity of the system. Thus, 
they are often viewed as an inexpensive alternative to building new transmission infrastructure \cite{pserc}. 
However, %increased penetration of 
increasing the number of 
RAS increases the operational complexity and poses several challenges as the power system continues to evolve rapidly. Existing procedures for designing RAS are often slow, requiring numerous offline simulations to ensure that the proposed action is sufficient and does not interact adversely with existing RAS and other protection devices \cite{pserc}, \cite{ras_failure}. 
As a result, parameters of RAS, such as the conditions specified to trigger the RAS and the type of actions taken, typically do not change during real-time operations \cite{pnnl}. The slow design procedure, coupled with the use of fixed control parameters, may prevent RAS from adapting to rapidly evolving grid conditions.

Recent research has sought to address some of the issues and risks associated with corrective action identification.
To improve the RAS design procedure, \cite{automation_ras} proposes a sensitivity-based method to generate a set of triggering conditions and generator tripping actions to address post-contingency line overloads. However, the proposed method manually identifies suitable RAS actions, which limits the number of contingencies, operating conditions and control actions that can be studied. 
% This method can be extended to use optimization methods to identify an optimal set of corrective actions, as is proposed in this paper. 
% \textcolor{red}{shorten paragraphs 4, 5, and 6}
%Additionally, RAS can also increase the risk of operation of power systems. 
 %Undesirable operation of RAS and interactions between multiple RAS can have severe consequences and result in widespread blackouts \cite{pserc},\cite{ras_failure}. 
 %
 %As undesirable operation of RAS and interactions between multiple RAS can have severe consequences and result in widespread blackouts \cite{pserc}, \cite{ras_failure}, the risk of RAS misoperations should be assessed. 
Further, recent research has shown the utility of considering the risk of corrective action failure (i.e. the risk of post-contingency corrective actions not being implemented correctly) in identifying an optimal system dispatch \cite{whither},\cite{rmac},\cite{whither-ac}. These methods considers corrective generator re-dispatch as the only corrective action, which simplifies modelling and solution of the optimization problem. However, generator re-dispatch
 is typically neither automatic nor quick enough to be considered as a RAS.

In this paper, we extend the above studies by proposing an optimization framework to design RAS and develop a cascading failure simulation to assess risk of RAS misoperation. % to assess potential risks of RAS misoperation. %Our work extends \cite{automation_ras} 

Our first contribution is an extension of the traditional security constrained optimal power flow (SCOPF) to design and optimize RAS settings at an operational time frame, which we refer to as RAS-SCOPF. The RAS-SCOPF is a mixed-integer optimization problem which models system operations in three stages, namely (i) pre-contingency operation, (ii) intermediate (post-contingency, pre-RAS) operations, and (iii) post-contingency, post-RAS operations. An innovative aspect of this model is that the second stage includes a set of logical constraints that describes whether or not the RAS is triggered, with RAS actions as decision variables.
%The RAS parameters, modelled as decision variables, include the triggering threshold (i.e., level of line overloads) that activates the RAS and the RAS actions (i.e., which generators are tripped once the RAS is triggered). The RAS parameters are shared among all considered contingencies (i.e., we assume that the RAS only relies on local line measurements rather than which contingency caused the overload). 

%The implementation of RAS allows us to relax the standard N-1 security constraints, resulting in a lower operational cost but also increased risk. 
Our second contribution is to develop an cascading failure simulation to assess the risk of the RAS not working as intended due to, e.g., unexpected operating conditions. %Due to its low computational complexity, a DC power flow based cascade simulator was implemented in Julia. 
Our cascading failure simulator is based on the DCSIMSEP simulator \cite{phines12}, but was reimplemented in Julia \cite{bezanson2017julia} and extended to include a model of the relevant RAS schemes.
%(which are not considered in DCSIMSEP).

% Instead of directly modelling the effects of RAS failure in the optimization model (as was done in \cite{whither}), we evaluate the impacts of RAS failure by performing cascading failure simulations (CFS) on the original test system without the RAS. To run the cascade simulations, we implemented a simulator, similar to DCSIMSEP \cite{phines12}, in Julia.  

% 

Our third contribution is to demonstrate our proposed method in a case study on the IEEE RTS-96 single area system. First, we show that the proposed RAS-SCOPF method reduce both operating cost and cascading risk compared to the traditional OPF and SCOPF formulations. %, including an optimal power flow without , SCOPF RAS-aware SCOPF. 
%in terms of their pre-contingency operating costs and their risk of post-contingency cascading failure. % using the cascading simulations. 
Second, we assess the risk of RAS misoperation under loading conditions different from what it was designed for. The results highlight the benefits of reoptimizing the RAS settings in operations.  
%We find that explicitly optimizing over RAS parameters allows us to reduce both operating cost and cascading risk.

% as the loading conditions
The rest of the paper is structured as follows. Section \ref{sec:OPF} describes the mathematical formulation of the proposed optimization problem, while Section \ref{sec:CFS} discusses the set-up of the cascading simulations. Section \ref{sec:case-study} presents the results of the case study, and Section \ref{sec:conc} concludes the paper.

\section{Optimal Design of Remedial Action Schemes} \label{sec:OPF}
%We next present the formulation of the RAS-SCOPF, before introducing the benchmark formulations. 
RAS schemes are typically used to mitigate the impact of particular contingencies on system operations, thus allowing more effective use of existing transmission capacity in normal operations. While RAS may resolve different kinds of post-contingency problems and may involve different kinds of control actions, we focus on alleviating post-contingency line overloads using generation tripping. Our optimization aims to optimally choose the RAS actions, i.e. which generators are tripped once the RAS is triggered. We assume that the RAS is triggered when the considered line is overloaded, %the real-time measurements of a line flows crosses the threshold, 
regardless of what caused the line flow to exceed the limit. As a result, the RAS action is shared among all considered contingencies. 

% (i) the triggering threshold, i.e. level of line overload, that activates the RAS and (ii)

\subsection{Formulation of the RAS-SCOPF}
%In this section we present the mathematical formulation of the proposed optimal power flow problem for designing remedial action schemes.
%{\color{red}We need a short paragraph to introduce the notation! What do the different sets represent? What are the subscripts and superscripts?}
%\subsubsection*{Notation}
We consider a power system where the sets $\mathcal{G},\mathcal{B}$ and $\mathcal{L}$ represent the generators, buses and lines in the system, and $|\mathcal{G}|$ represents the number of elements in $\mathcal{G}$. 
The parameters $P_{g}^{max}$ and $P_g^{min}$ represent the maximum and minimum generation limits, while $P_f^{max}$ is the maximum transmission line capacity.
%Variables and parameters associated with the pre-contingency, post-contingency and post-RAS operating stages are denoted by superscripts $^o$, $^i$ and $^c$, respectively.
We use \textbf{bold fonts} for decision variables, with the vectors $\boldsymbol{P}_{\boldsymbol{g}}^{\boldsymbol{o}}$, $\boldsymbol{\theta^o}$ and $\boldsymbol{P_f^o}$ representing the pre-contingency generation, voltage angles and power flows, respectively. Similar decision variables with superscripts $i$ and $c$ are used to represent the power flows in the intermediate and the post-RAS stage, respectively. Generator and voltage variables related to individual generator or buses are denoted with subscript $i$ or $j$, e.g. $\boldsymbol{P}_{\boldsymbol{gi}}^{\boldsymbol{o}}$, while lines have double subscripts $ij$ representing either ends of the line, e.g. $\boldsymbol{P}_{\boldsymbol{fij}}^{\boldsymbol{o}}$.
%The parameters $P_d^o$ represents the loads, while $B^F$ represents line admittance matrix. The sets $\mathcal{G},\mathcal{B}$ and $\mathcal{L}$ represent the generators, buses and lines in the system. %The parameters $P_f^{max}$, $P_{g}^{max}$ and $P_g^{min}$ represent the maximum transmission capacity of the lines, and the maximum and minimum generation limits, and  respectively. 
% Decision variables with superscripts $o$, $i$ and $c$ are used to represent the power flows in the pre-contingency, intermediate and post-RAS stage, respectively.

\subsubsection{Objective function} The objective function is given by
\begin{equation}
    \label{eq:obj}
    \min ~\sum_{i\in\mathcal{G}} f_i(\boldsymbol{P_{gi}^o})
    +\!\sum_{k \in \mathcal{C}} \!\left(\sum_{i\in\mathcal{B}}\gamma(\boldsymbol{P_{di}^{c,k}}\!-\!P_{di}^o)
    +\!\!\sum_{i \in \mathcal{G}} \rho(1\!-\!\boldsymbol{z_{gi}^j})\!\right)
\end{equation}
The first term represents the generation cost of normal operation, with %with cost per generator given by
%. The generation cost per generator is given by
$$f_i(\boldsymbol{P_{gi}^o})=c_{2,i}(\boldsymbol{P_{gi}^o})^2 + c_{1,i}\boldsymbol{P_{gi}^o}$$
where $c_{2,i}$ and $c_{1,i}$ are the quadratic and linear cost coefficients of generator $i$, respectively.
The second term %is the impact of system failure on the user. In our case, the impact of system failure is 
penalizes post-contingency load shedding, with $P_{di}^o$ representing the normal load and $\boldsymbol{P_{di}^{c,k}}$ representing load served after a RAS is triggered. The parameter $\gamma$ represents the load shedding cost. The third term minimizes the magnitude of the RAS action, represented as the number of generators tripped multiplied by a penalty factor $\rho$. 

\subsubsection{Pre-contingency operating constraints}
We use a DC power flow model to model our system. %, with the pre-contingency generation $\boldsymbol{P}_{\boldsymbol{g}}^o$, voltage angles $\theta^o$ and power flows $\boldsymbol{P}_{\boldsymbol{g}}^o$ as decision variables. 
The pre-contingency nodal power balance and power flows are given by %, with \eqref{eq:prc_dc} representing the pre-contingency power flows.
\begin{align}
    & \boldsymbol{P_{gi}^o}-P_{di}^o=\sum_{j \in \mathcal{B}}\boldsymbol{P_{fij}^o}  \quad &&\forall i \in \mathcal{B}\label{eq:prc_nb}\\
    & \boldsymbol{P_{fij}^o}= -b_{ij}(\boldsymbol{\theta_i^o}-\boldsymbol{\theta_j^o}) && \forall ij\in\mathcal{L}
\end{align}
where $b_{ij}$ is the admittance of line $ij$.
% with the power flows defined as 
% \begin{align}
%     P_{fij}^o= -B_{ij}(\theta_i^o-\theta_j^o)\quad \forall ij \in \mathcal{L}. \label{eq:prc_dc}
% \end{align}
Generator and line limits are enforced by 
\begin{align}
    & P_{gi}^{min} \leq \boldsymbol{P_{gi}^o} \leq P_{gi}^{max} &&\forall i \in\mathcal{G} \label{eq:prc_gl}\\
    -&P_{fij}^{max} \leq \boldsymbol{P_{fij}^o} \leq P_{fij}^{max} &&\forall ij\in\mathcal{L} \label{eq:prc_ll}
\end{align}

\subsubsection{Intermediate operating constraints}
We next model the intermediate operating condition just after each contingency (prior to the implementation of any RAS). These constraints are included for all contingencies $k \in \mathcal{C}$, where $\mathcal{C}$ is the set of all contingencies. The subset of critical contingencies that the RAS is designed to protect against is denoted by $\mathcal{C}_M \subset \mathcal{C}$. % that result in overloads in other lines in the system.

To make up for generation imbalances following a generation or load outage, we assume a distributed slack model and redispatch generators using pre-determined participation factors $K_i$. The new generation levels $\boldsymbol{P_g^{i,k}}$ and associated generation limit constraints are given by 
\begin{align}
    &\boldsymbol{P_{gi}^{i,k}} \!=\! \boldsymbol{P_{gi}^o} + K_i\left[\sum_{i\in\mathcal{B}}(P_{di}^{i,k}\!-\!P_{di}^o) +\sum_{i\in\mathcal{G}}(\boldsymbol{P_{gi}^o}\!-\!\boldsymbol{P_{gi}^{i,k}}) + \boldsymbol{\Delta}^i \right]\!, \nonumber \\ 
    & \qquad\qquad\qquad\qquad\qquad\qquad\qquad\qquad\quad \forall i\in\mathcal{G}_k, k \in \mathcal{C} \label{eq:poc_rd}
    \\
    &P_{gi}^{min} \leq \boldsymbol{P_{gi}^{i,k}} \leq P_{gi}^{max}, \qquad\qquad\qquad~~~\forall i\in\mathcal{G}, k \in \mathcal{C} \label{eq:poc_gl}
\end{align}
where $\mathcal{G}_k$ is the set of online generators after contingency $k$ and the variable $\boldsymbol{\Delta}^i$ represents the power mismatch in the network that arises because the participation factors $K_i$ of the online generators may not sum to 1. % (due to e.g. generator outages). 
% It is used to address the fact that after generators are shed, the participation factors of the remaining generators in the system do not add up to one. 
%{\color{red}where $\Delta^i$ represent...}
The intermediate power balance and power flow constraints are given by 
\begin{align}
    %& \text{for all  $i \in \mathcal{B}$, contingencies $k \in \mathcal{C}$} \nonumber\\
    & \boldsymbol{P_{gi}^{i,k}}-P_{di}^{i,k} = \sum_{j \in \mathcal{B}} \boldsymbol{P_{fij}^{i,k}} && \forall i \in \mathcal{B}, k \in \mathcal{C}\label{eq:poc_nb} \\
    %& \text{for all  $(i,j) \in \mathcal{L}$, contingencies $k \in \mathcal{C}$} \nonumber\\
    & \boldsymbol{P_{fij}^{i,k}}= -b_{ij}(\boldsymbol{\theta_{i,k}^i}-\boldsymbol{\theta_{j,k}^i}) &&\forall ij \in \mathcal{L}_k, k \in \mathcal{C} \label{eq:poc_dc}
\end{align}
where $\mathcal{L}_k$ is the set of non-outaged lines in contingency $k$. For the contingencies the RAS is designed to protect against, denoted by $\mathcal{C}_M$, we do not enforce power flow limits, as we assume there may be overloads. For all other contingencies $k\in\mathcal{C}\backslash\mathcal{C}_M$, we enforce post-contingency line limits,
\begin{align}
    &-P_{fij}^{max} \leq \boldsymbol{P_{fij}^{i,k}} \leq P_{fij}^{max} && \forall ij \in \mathcal{L}, k \in \mathcal{C}\backslash\mathcal{C}_M \label{eq:poc_ll}
\end{align}
%, which differs from the pre-contingency line admittance matrix  $B^{F}$ in the case of line outages.
\subsubsection{RAS design and triggering constraints}
In the intermediate stage, we also include constraints to evaluate whether the RAS is triggered by an overload. These constraints are included for all lines in the set of monitored lines, denoted by $\mathcal{L}_M$, and for the contingencies $\mathcal{C}_M$. Note that the set $\mathcal{L}_M$ can include one or more lines. 
%
%The idea  behind the modelled RAS is that 
The following logic constraints assess whether the loading on a monitored line $ij \in \mathcal{L}_M$ exceeds its maximum loading after a contingency $k$,
\begin{align}
    \boldsymbol{P_{fij}^{i,k}}-P_{fij}^{max}  &\geq m(1-\boldsymbol{z_{1,ij}^k}) && \forall ij\in\mathcal{L}_M,k\in\mathcal{C_M}\label{eq:pos_no_ol}\\
    \boldsymbol{P_{fij}^{i,k}}-P_{fij}^{max} &\leq M\boldsymbol{z_{1,ij}^k} && \forall ij\in\mathcal{L}_M,k\in\mathcal{C_M}\label{eq:pos_ol}\\
    -\boldsymbol{P_{fij}^{i,k}}-P_{fij}^{max} &\geq m(1-\boldsymbol{z_{2,ij}^k}) && \forall ij\in\mathcal{L}_M,k\in\mathcal{C_M}\label{eq:neg_no_ol}\\
    -\boldsymbol{P_{fij}^{i,k}}-P_{fij}^{max} &\leq M\boldsymbol{z_{2,ij}^k} && \forall ij\in\mathcal{L}_M,k\in\mathcal{C_M}\label{eq:neg_ol}\\
    \boldsymbol{z_{1,ij}^k} + \boldsymbol{z_{2,ij}^k} &\geq \boldsymbol{z_{3,ij}^k} && \forall ij\in\mathcal{L}_M,k\in\mathcal{C_M}\label{eq:no_ol} \\
    \boldsymbol{z_{1,ij}^k} + \boldsymbol{z_{2,ij}^k} &\leq \boldsymbol{z_{3,ij}^k} && \forall ij\in\mathcal{L}_M,k\in\mathcal{C_M}\label{eq:ol}
\end{align}
% exceeds the maximum loading by more than the triggering threshold $\lambda_{RAS}$
Eqs. \eqref{eq:pos_no_ol} and \eqref{eq:pos_ol} set the variable $\boldsymbol{z_{1,ij}^k}=1$ if the line is overloaded in the positive flow direction, and $\boldsymbol{z_{1,ij}^k}=0$ otherwise. Eqs. \eqref{eq:neg_no_ol}, \eqref{eq:neg_ol} and $\boldsymbol{z_{2,ij}^k}\in\{0,1\}\}$ enforce the same condition for the negative flow direction. If the line is overloaded in either direction, \eqref{eq:no_ol} and \eqref{eq:ol} set $\boldsymbol{z_{3,ij}^k}=1$ to indicate that contingency $k$ causes a RAS-triggering condition on line $ij$, and ensure that $\boldsymbol{z_{3,ij}^k=0}$ otherwise. The parameters $M$ and $m$ are big-M constants that represent valid lower and upper bounds on the left hand side of the constraints. 

The binary variable $\boldsymbol{y_{j}^k}$ indicates whether the $j^{th}$ RAS scheme has been triggered. Specifically, $\boldsymbol{y_{j}^k}=1$ if $\boldsymbol{z_{3,ij}^k}=1$ for one or more lines in the monitored set $\mathcal{L}_M^j$ and $y_{j}^k=0$ otherwise. For all $ k\in\mathcal{C}_M$, this condition is expressed by
\begin{align}
    % & \sum_{(i,j) \in \mathcal{L}_M} \boldsymbol{z_{3,ij}^k} \geq \boldsymbol{y_{j}^k} \label{eq:no_trig}\\
    % & \sum_{(i,j) \in \mathcal{L}_M} \boldsymbol{z_3^{ij}} \leq |\mathcal{L}_M|\boldsymbol{y_{j}^k}  \label{eq:trig}
    & \sum_{(i,j) \in \mathcal{L}_M^j} \!\!\!\!\boldsymbol{z_{3,ij}^k} \geq \boldsymbol{y_{j}^k} \qquad \sum_{(i,j) \in \mathcal{L}_M^j} \!\!\!\!\boldsymbol{z_3^{ij}} \leq |\mathcal{L}_M|\boldsymbol{y_{j}^k}  \label{eq:trig}
\end{align}
%In \eqref{eq:trig}, $M$ is the number of monitored lines. 
For the $j^{th}$ RAS scheme, we assume that at least one generator should be tripped, i.e., 
\begin{align}
    \sum_{i\in\mathcal{G}} \boldsymbol{z_{gi}^j}\leq |\mathcal{G}| - 1 \label{eq:gen_trip}
\end{align}
where $\boldsymbol{z_{gi}^j}$ represents the post-RAS status of generator $i\in\mathcal{G}$. If the generator is still operating, $\boldsymbol{z_{gi}^j} = 1$ and otherwise $\boldsymbol{z_{gi}^j}=0$. 
%where $y_{j}^k=1$ implies that the $j^{th}$ RAS has been triggered.
% {\color{red}where $zy{j}^k=1$ indicates...}

%{\color{red}How do these constraints ensure that the RAS action remains the same for all contingencies?}

\subsubsection{Post-RAS constraints}
%We next model the post-RAS operating condition. %, i.e. how the system will operate if $\boldsymbol{y_{j}^k}=1$ for any of the considered RAS schemes.
For all contingencies $k \in \mathcal{C}_M$, if $\boldsymbol{y_{j}^k}=1$ and the $j^{th}$ RAS is triggered after contingency $k$, we enforce that the generators are tripped as described by \eqref{eq:gen_trip}, 
\begin{align}
    % & \boldsymbol{z_{gi}^k} - \boldsymbol{z_{gi}^j} \leq (1-\boldsymbol{y_j^k})\\
    % & \boldsymbol{z_{gi}^k} - \boldsymbol{z_{gi}^j} \geq (\boldsymbol{y_j^k}-1)
    & \boldsymbol{z_{gi}^k} - \boldsymbol{z_{gi}^j} \leq (1-\boldsymbol{y_j^k}) \qquad
    \boldsymbol{z_{gi}^k} - \boldsymbol{z_{gi}^j} \geq (\boldsymbol{y_j^k}-1)
\end{align}
If the RAS is not tripped, i.e. $\boldsymbol{y_{j}^k}=0$, all generators will remain in operation,
\begin{align}
    & \sum_{i\in\mathcal{G}} \boldsymbol{z_{gi}^k} \geq|\mathcal{G}|(1-\boldsymbol{y_{j}^k})\label{eq:no_gen_trip}
    % This constraint forces all z_gi's to be 1 id y_j^k=0
    %
    % & \sum_{i\in\mathcal{G}} z_{gi}-N_g + 1 \leq M(1-y_{j}^k) \label{eq:gen_trip}
    % & \sum_{i\in\mathcal{G}} z_{gi}^j\leq N_g - 1\label{eq:gen_trip}\\
    % This constaint forces at least one z_{gi}^j=0 (i.e. at least one generator is tripped in the j^th RAS)
    %
    % These two constraints force z_{gi}^k = z_{gi}^j if if y_j^k = 1 (otherwise free)
\end{align}
%Here, $N_g$ is the number of generators in the system.  %As before $m$ and $M$ are big-M constants representing valid upper and lower bounds on the left-hand side of the constraints.
% {\color{red}Here, $z_{gi}$ is a variable that represent the post-RAS status of generator $i\in\mathcal{G}$, with $z_{gi}=1$ if the generator is online and $z_{gi}=0$ if the generator is tripped. 
% Say what m and M are, what N and g (or maybe it should be $N_g$?) represents...}
In addition to generator tripping, we also allow load shedding as an emergency action, i.e.,
\begin{align}
    % & 0 \leq \boldsymbol{P_{di}^{c,k}} \leq P_{di}^{i,k} \qquad \forall i\in\mathcal{B}\label{eq:load_shed}
    & P_{di}^{i,k}(1-\boldsymbol{y_{j}^k}) \leq \boldsymbol{P_{di}^{c,k}} \leq P_{di}^{i,k} \qquad \forall i\in\mathcal{B}\label{eq:load_shed}
\end{align}
%{\color{red}Why is only the sum constrained to be larger? How about for individual loads?}
The load shedding may differ between contingencies, but is penalized with a very high value $\gamma$ in the objective function. 

%To represent the adjustment in generation needed to manage the imbalance following the generation trip triggered by the RAS. For contingencies not to be isolated by the RAS ($\mathcal{C}\backslash\mathcal{C}_M$),  it is assumed that the generators and loads remain fixed at their values before RAS action.

% \eqref{eq:no_rd}-\eqref{eq:no_ls}. Equations \eqref{eq:por_dc}-\eqref{eq:por_ll} represent the power flow in this post-RAS conditions.

%Constraints \eqref{eq:gen_trip}-\eqref{eq:por_ll} represent the post-contingency operating conditions after the RAS has been implemented (if the overload is high enough). Equations \eqref{eq:gen_trip} and \eqref{eq:no_gen_trip} indicate that if the RAS is triggered, we may choose to trip a subset $\mathcal{G}_{RAS}$ of the set of generators $\mathcal{G}$. In situations where generator tripping alone is not sufficient to alleviate the overloads in the monitored lines, \eqref{eq:load_shed} facilitates load shedding as an emergency action. 

To manage the imbalance following the generation trip triggered by the RAS as well as load shedding, we again use adjustments based on droop factors $K_i$, i.e.
\begin{align}
    & \boldsymbol{P_{gi}^{c,k}} \!=\! \boldsymbol{P_{gi}^{i,k}} \!+\! K_i\! \left[\sum_{i\in\mathcal{B}}(\boldsymbol{P_{di}^{c,k}}\!-\!P_{di}^{i,k}) +\!\! \sum_{i\in\mathcal{G}}(\boldsymbol{P_{gi}^{i,k}}\!-\!\boldsymbol{P_{gi}^{c,k}}) \!+\! \Delta^c \right]
    \label{eq:por_rd}
\end{align} 
We enforce that the generation levels either respect the generation limits or are set to zero if the generator is tripped,
\begin{align}
    &  \boldsymbol{z_{gi}^k}P_{gi}^{min} \leq \boldsymbol{P_{gi}^{c,k}} \leq \boldsymbol{z_{gi}^k}P_{gi}^{max} \label{eq:por_gl}
\end{align}
% For contingencies $k\in\mathcal{C}\backslash\mathcal{C}_M$ that are not considered to be mitigated by RAS (i.e., are not allowed to cause post-contingency overloads and thus will not trigger any RAS scheme), we enforce that the generators and loads remain fixed at their values before RAS action,
% \begin{align}
%     & \boldsymbol{P_{gi}^{c,k}} = \boldsymbol{P_{gi}^{i,k}}
%     \qquad
%     %\label{eq:no_rd}\\
%     \boldsymbol{P_{di}^{c,k}} = P_{di}^{i,k} \label{eq:no_ls}
% \end{align}
The power flow in the post-RAS conditions is represented by 
\begin{align}
    & \boldsymbol{P_{gi}^{c,k}}-\boldsymbol{P_{di}^{c,k}}=\sum_{j \in \mathcal{B}} \boldsymbol{P_{fij}^{c,k}} \label{eq:por_nb}\\
    & \boldsymbol{P_{fij}^{c,k}}= -b_{ij}(\boldsymbol{\theta_{i,k}^c}-\boldsymbol{\theta_{j,k}^c}) \label{eq:por_dc} \\
    &-P_{fij}^{max} \leq \boldsymbol{P_{fij}^{c,k}} \leq P_{fij}^{max}\label{eq:por_ll}
\end{align}

%\subsubsection{Full RAS-SCOPF problem} 
To summarize, the RAS-SCOPF minimizes the pre-contingency operation cost and chooses the smallest RAS action that reduces load shedding and maintains secure operation against a set of specified contingencies.
% The full RAS-SCOPF problem can be summarized as 
% \begin{align}
%     &\min &&\text{Objective } \eqref{eq:obj} \nonumber \\
%     &~\text{s.t.} && \text{Constraints }\eqref{eq:prc_nb}-\eqref{eq:por_ll} \nonumber
% \end{align}
%In constraints \eqref{eq:poc_rd} and \eqref{eq:por_rd}, generator re-dispatch is modelled using a distributed slack bus model to capture the behaviour of fast-acting primary controls of the generators. Constraints \eqref{eq:pos_no_ol}-\eqref{eq:no_gen_trip} use of binary decision variables to model the triggering of RAS and the choice of generators to be tripped. 
Since the pre-contingency operation cost is a quadratic function of the pre-contingency generation, the resulting problem is a mixed integer quadratic program (MIQP).%As a result, the problem does not scale well as its size increases. For larger systems and a bigger set of load scenarios and contingencies, the resulting formulation may not be tractable. 

\subsection{Benchmark formulations}
To assess the performance of the RAS-SCOPF, we compare its performance against a traditional DC OPF and a DC SCOPF, as described below. \\ %We also introduce a modified version of the -OPF that incorporate security constraints for the contingencies not considered in the RAS design. 
%These formulations can be defined as modifications of the formulation above:\\
% formulation proposed in section \ref{sec:OPF}
%\begin{enumerate}
    %\item 
    \textit{1) OPF}: The OPF minimizes pre-contingency generation cost, 
    \begin{equation}
        \min \sum_{i\in\mathcal{G}} f_i(\boldsymbol{P_{gi}^o})
        \label{eq:pre-cont-cost}
    \end{equation}
    subject only to the pre-contingency constraints \eqref{eq:prc_nb}-\eqref{eq:prc_ll}. \\%  i.e.,
    % \begin{align}
    %     & \min && \sum_{i\in\mathcal{G}} f_i(\boldsymbol{P_{gi}^o}) \\
    %     & \text{~s.t.} &* Pre-contingency constraints \eqref{eq:prc_nb}-\eqref{eq:prc_gl}
    % \end{align}
    %To solve the OPF, we minimize the pre-contingency generation cost subject only to the pre-contingency constraints \eqref{eq:prc_nb}-\eqref{eq:prc_gl}.
    %\item 
    \textit{2) SCOPF}: The SCOPF minimizes the pre-contingency generation cost \eqref{eq:pre-cont-cost} subject to the pre-contingency constraints \eqref{eq:prc_nb}-\eqref{eq:prc_ll} and security constraints \eqref{eq:poc_rd}-\eqref{eq:poc_ll}. to ensure that the system will operate without violations immediately after any contingency has occurred,
    %\eqref{eq:poc_rd}-\eqref{eq:poc_dc}. 
    %The contingency set $\mathcal{C}$ only includes line outages on non-radial lines and do not lead to any power imbalances in the system, 
    and
    we set $\mathcal{C}_M=0$.
    %We thus enforce that the generator outputs remain the same in the post-contingency state, i.e. we set $P_{gi}^{i,k}=P_{gi}^o$. %Hence, we replace the constraints \eqref{eq:poc_rd} and \eqref{eq:poc_gl} with the condition $P_{gi}^{i,k}=P_{gi}^o$ which ensures that the generation remains the same in the post-contingency condition.  %corresponds to the re-dispatch of generators when a contingency occurs. 
    % The post-contingency power flow is modelled by constraints \eqref{eq:poc_nb}, \eqref{eq:poc_dc} and we impose post-contingency line limits, i.e., % the line limits in the post-contingency state to ensure that the system remains secure after an N-1 contingency occurs, i.e.,
    % \begin{equation}
    %     -P_{fij}^{max}\leq P_{fij}^{i,k} \leq P_{fij}^{max}
    % \end{equation}
    %\item 
    
    \noindent\textit{3) RAS-aware SCOPF}: The RAS-aware SCOPF is a modified version of the SCOPF that does not include the normal security constraints for contingencies $k\in\mathcal{C}_M$. 
    The RAS-aware SCOPF minimizes pre-contingency generation cost \eqref{eq:pre-cont-cost} subject to the pre-contingency constraints \eqref{eq:prc_nb}-\eqref{eq:prc_ll} and security constraints \eqref{eq:poc_rd}-\eqref{eq:poc_ll} for $k\in\mathcal{C}\backslash\mathcal{C}_M$. For contingencies $k\in\mathcal{C}_M$, the RAS-aware SCOPF ensures feasibility of post-RAS generation redispatch by enforcing
    % security constraints for all contingencies $\mathcal{C}\backslash\mathcal{C}_M$. For any contingency in $\mathcal{C}_M$, the feasibility of post-RAS generation redispatch is ensured by enforcing the following constraints
    \begin{align}  
            & \boldsymbol{P_{gi}^o} + \boldsymbol{r_i} \leq P_{gi}^{max} &&\forall i\in \mathcal{G}, k\in\mathcal{C}_M\\
            & \boldsymbol{r_i} \geq K_i \Delta P_g &&\forall i\in \mathcal{G}, k\in\mathcal{C}_M\\
        & \Delta P_g = \sum_{i \in \mathcal{G}_{RAS}} \boldsymbol{P_{gi}^o} &&\forall  k\in\mathcal{C}_M
    \end{align} 
    where $\boldsymbol{r_i}$ is a reserve capacity needed to handle RAS activation and $\mathcal{G}_{RAS}$ is the set of generators tripped by the RAS. %Also, since the RAS is designed to protect the system against all contingencies in the set $\mathcal{N}_c$, the corresponding constraints are relaxed.
%\end{enumerate}

% RAS-aware SCOPF purpose-RAS designed only for one load scenario. When simulating for other load scenarios, need to initialize the system appropriately. For the new loading conditions, dispatch should be such that the RAS actions are feasible-generator redispatch should be feasible. Thus we have to wnusure enough reserve

% At the same time, we expect the RAS to prevent overloads in the event of contingencies that were considered during it design. Thus, all constraints corresponding to these contingencies are relaxed. 

% When designing the RAS, only a subset of all N-1 contingencies, $\mathcal{N}_c$, was considered. Therefore, any other N-1 contingencies could result in a blackout. To ensure that the system is secure against all other contingencies, we use the solutions of the RAS-aware SCOPF to dispatch all the generators. The RAS-aware SCOPF is a modified version of the SCOPF

% While designing the RAS, only a single load scenario is considered. Thus, when evaluating the RAS for different load conditions, the system needs to initialized appropriately. Thus, we use the RAS-aware SCOPF, which is a modified SCOPF, to dispatch the system under varying loading conditions. For new loading conditions, the generators have to dispatched such that after the RAS action, the generator redispatch is feasible. This is ensured by adding the following constraints,

\section{Cascading simulation} \label{sec:CFS}
% \begin{itemize}
% \item DCPF based-computationally efficient, however has limitations. steady state assumption after every contingency, may not be valid after the first few as the system nears instability, by choosing appropriate definition of system failure, can terminate the algorithm before the system becomes unstable. 
% \item Dynamic model-more accurate representation of system evolution, however can be very complicated, require meticulous calibration, and computationally intensive. 
% \end{itemize}
% We rely on the DCPF based CFS becuase we wish to evaluate the RAS's abillity to arrest cascading failure. If the RAS fails in doing so first few iterations of the CFS, we can conclude that it is not sufficient. 

Modelling cascading failures in power systems is challenging because there are several ways in which an initiating contingency can evolve into a series of cascading outages, e.g., cascading thermal overloads, voltage instability, transient instability, hidden failures in protection systems or human errors \cite{cf_risk},\cite{cf_survey}.
%
% Different model are capable of capturing different aspects of the cascading. Model based on DC power flow are computationally very efficient. Despite their advantages, these models are limited in their application because of the assumption that the system reaches a steady state. As a result, these models capture the full behaviour of the system under conditions of transient instability, voltage instability, etc. Even when studying steady-state phenomena like cascading thermal overloads, the assumption that the system reaches a steady-state after a contingency fails after the first few contingencies. 
%
%Different models are capable of capturing different aspects of cascading failures. 
Here, we focus on cascading events driven by thermal overloads, as our goal is to evaluate the effectiveness of a RAS scheme in preventing such cascading events.
%Hence, we focus on cascading events driven by thermal overloads and use a DC power flow based cascading model.
We base our cascading simulator on a DC power flow model, which is computationally very efficient and hence widely used to assess system behaviour when subjected to multiple contingencies and operating conditions. %Despite their advantages, the DC power flow based models assume that system reaches a steady-state after every contingency and thus, cannot capture the effects of dynamic phenomena like voltage and transient instability. Even when studying steady-state phenomena, like cascading thermal overloads, the existence of a steady-state is not guaranteed after the first few contingencies. 
The drawback of using a DC power flow based model is that it does not capture reactive power and voltage variability, and also assumes that the system reaches a steady-state after every contingency. Thus, DC models cannot capture the effects of dynamic phenomena like voltage and transient instability, or in situations where a steady-state solution does not exist.
However, in the early stages of a cascade, before the loss of dynamic stability, the DC power flow models can describe the evolution of the system with good accuracy.
Thus, choosing an appropriate definition of system failure helps limit the difference between actual system behaviour and that predicted by models based on DC power flow approximations in the case of cascading overloads 
\cite{compare_cfs}. 
%
%Dynamic transient stability models, on the other hand,  offer a more accurate representation of the system's behaviour during sequence of cascading outages. However, they are computationally intensive and complicated, often needing meticulous tuning. Even a full dynamic model is not completely accurate because many system parameters are not exactly known. 
%
%Our objective is to evaluate the designed RAS's ability to prevent cascading line overloads. 
%If we reach system failure before the RAS is triggered, then we can conclude that the RAS design was not adequate to prevent the spread of the contingency. %In the early stages of the simulation, before the loss of dynamic stability, the DC power flow models can describe the evolution of the system with good accuracy. Hence, we used a simulator based on DC power flow approximation to evaluate the performance the designed RAS. 
% (building on the one presented in \cite{phines12})

\begin{figure}
    \centering
    \includegraphics[width = 0.7\columnwidth]{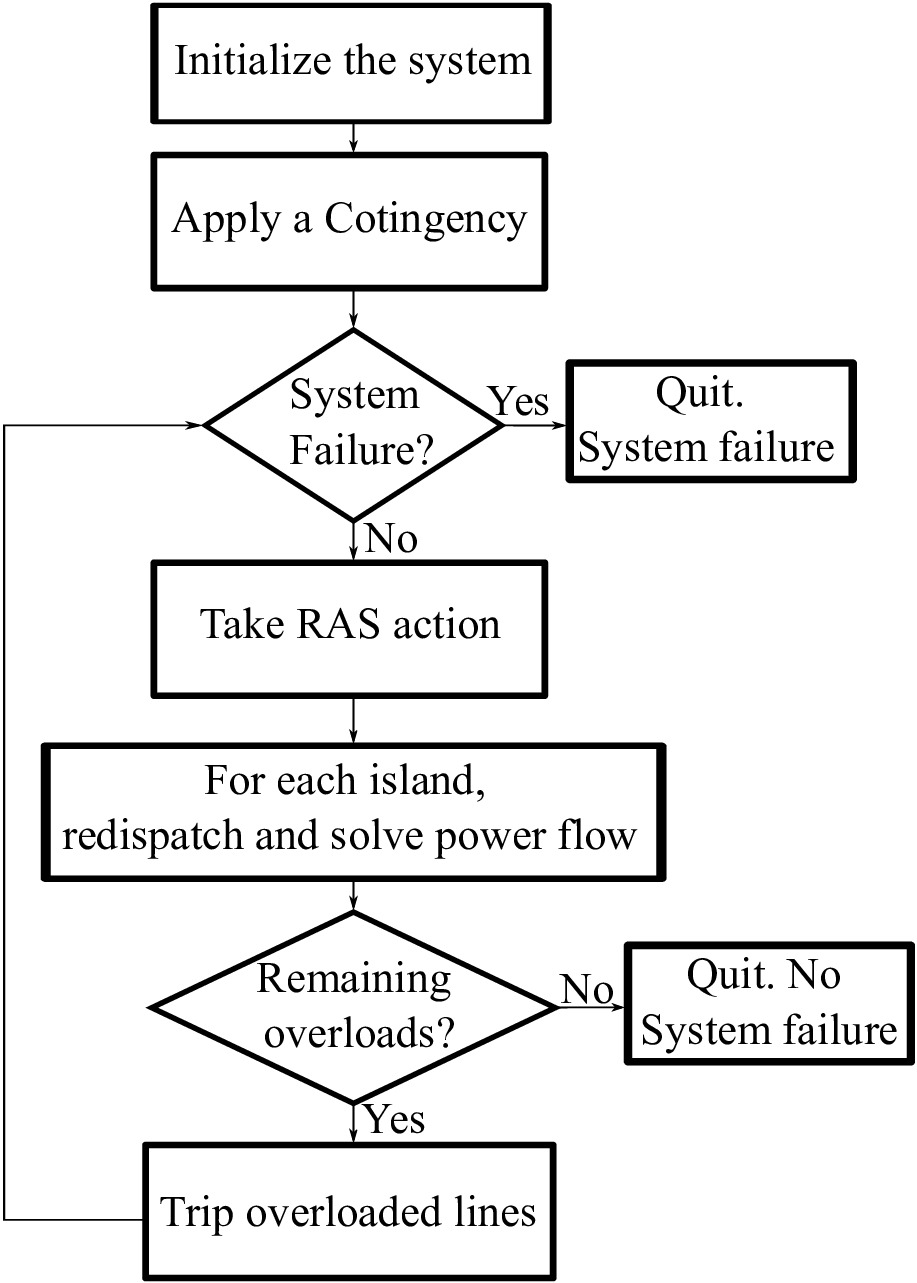}
    \caption{Flowchart for the Cascading Failure Simulator \vspace{-10mm}}
    \label{fig:cfs}
\end{figure}

Figure \ref{fig:cfs} illustrates the cascading failure simulator designed to test the performance of the designed RAS. Our simulator builds on DCSIMSEP \cite{phines12}. A major difference between DCSIMSEP and our simulator is RAS modelling, which is absent in DCSIMSEP.
Another difference includes the way in which generators and loads are redispatched, which we do by solving the optimization problem \eqref{eq:obj_rd}-\eqref{eq:ls_rd}. %to redispatch generators and loads. 
Further, our simulator does not model overcurrent relays to track the time elapsed between successive outages. 

The steps involved in a cascading simulation are summarized in Fig. \ref{fig:cfs} and described below in more detail:

% Our cascading failure simulator is similar to DCSIMSEP \cite{phines12},not  but differs in the way generators are redispatched. Our simulator uses an optimization model \eqref{eq:obj_rd}-\eqref{eq:ls_rd} to redispatch generation. Another important difference is in the way in which time between successive events is tracked. DCSIMSEP models overcurrent relays to track the time between successive line outages. Since our aim is to test the effectiveness of the RAS, we do not keep track of the time elapsed between successive outages.

\noindent \emph{\underline{Step 0: Initialization}} 
At the beginning of the simulation, the system has to be initialized appropriately. When evaluating the risk of cascading failures of the different formulations, the system is initialized using the RAS-SCOPF, OPF or SCOPF, respectively. When analysing the effectiveness of the RAS for different loading conditions, the system is initialized with the solutions of the RAS-aware SCOPF. % to ensure that the post-RAS generation redispatch is feasible under all loading conditions. 
% Using the solutions of the optimization problem described in section \ref{sec:OPF}, initialize the generation and transmission line flows in the system. When simulating the system without any RAS, initialize the system by solving a OPF. For cases with RAS, either use the solutions of section \ref{sec:OPF} or run a modified SCOPF to dispatch generators with enough reserve capacity. 

% Two cases - evaluating different formulations and their associated risk of cascading failure - initialize with OPF,RAS-SCOPF, SCOPF
% evaluating the sensitivity of the RAS to different loading conditions - RAS-aware SCOPF

% Before starting the simulation, initialize the system. Depending the purpose of the Cascading simulation, appropriate formualtio is used to initialize the simulation. When analyzing the evolution of the system without an RAS, the standard OPF is used. When analysing the effects of the introduction of RAS, we use the solutions of the RAS-SCOPF to initialize the system. Since, we are using only one load scenario to design the RAS, when analysing performance under different load conditions, we initialize the system by solving the RAS-aware SCOPF to ensure that the generators with non-zero participation factors $K_i$ have enough reserve.

\noindent \emph{\underline{Step 1: Apply a contingency}} We apply an $n-1$ contingency to the system by modifyng the status of the outaged line. 

\noindent \emph{\underline{Step 2: Check for system failure}} 
We calculate the line admittance matrix $B^F$ considering all outages in the system and identify all the resulting islands.
Here, we define system failure as the state when at least 10\% of the buses are disconnected from the largest island.
If the system satisfies the chosen definition of system failure, terminate the simulation. 
If it is not satisfied, continue to step 3.

\noindent \emph{\underline{Step 3: Implement RAS action}} 
If the applied contingency %triggers the RAS, i.e., it 
causes overloads in any one of the lines monitored by a RAS, this RAS is triggered. % The relevant generators with $z_{g,i}^j=0$ are tripped.  take the optimal action computed as the solution of the RAS-SCOPF. 
In this case, we trip generators according to the pre-defined RAS action and shed the necessary amount of load. If there are no RAS present in the system or if the RAS has already been triggered, we do nothing.
% If the applied contingency, triggers the RAS , i.e., causes overloads in any of the montiored lines, take the RAS action by changing the status of the actuated generators and shedding some load (from the solution of the opf problem in section \ref{sec:OPF}). If an RAS is not present in the system, skip this step. 

\begin{table*}%[t]
    \caption{Critical contingencies and Post-Contingency Loading Level on Overloaded Lines}
    \vspace{-2mm}
    \begin{center}  
    \begin{tabular}{l|ccccccccc}
    %{|c|c|c|c|c|c|c|c|c|c|}
    \hline
    & &  &  & & & & & & \\[-8pt]
        \textbf{Outaged Line} & 23 & 7 & 18 & 21 & 22 & 27 & 29 & 25 & 26 \\
        %\hline
        \textbf{Overloaded Line} & 7 & 23 & 23 & 23 & 23 & 23 & 23 & 28 & 28 \\
        \hline   
    & &  &  & & & & & & \\[-8pt]
        \textbf{Loading Level (\% of Max)} & 102.02\% & 120.37\% & 100.84\% & 108.33\% & 111.00\% & 120.37\% & 108.18\% & 103.99\% & 103.99\% \\
         \hline
    \end{tabular}  
    \vspace{-5mm}
    \end{center}
    \label{tab:cri_con}
\end{table*}

\noindent \emph{\underline{Step 4: Redispatch generators and load}} 
If there is a mismatch between load and generation because the system is separated into several islands or a RAS scheme has lead to generation tripping and load shed, we redispatch generators and loads in every island by solving the following optimization problem: %\eqref{eq:obj_rd}-\eqref{eq:ls_rd}. 
\begin{align}
    \centering
    & \min  \sum_{i\in\mathcal{B}_{isl}}\boldsymbol{P_{di}^n}-P_{di}^o + \sum_{i \in \mathcal{G}_{isl}} (1-\boldsymbol{z_{gi}^n})\label{eq:obj_rd} \\
    & \text{subject to} \nonumber \\
    & \text{for all generators in the island $i \in \mathcal{G}_{isl}$}\\
    & \boldsymbol{z_{gi}^n} \leq z_{gi}^o\\
    \begin{split} 
      & \boldsymbol{P_{gi}^n} -\Biggl(P_{gi}^o + K_i\,\sum_{i\in\mathcal{B}_{isl}}(\boldsymbol{P_{di}^n}-P_{di}^o) \\
      &  \qquad+ K_i\left[\sum_{i\in\mathcal{G}_{isl}}(\boldsymbol{P_{gi}^n}-P_{gi}^o) + \Delta \right]\Biggr) \!\leq\! M(1\!-\!\boldsymbol{z_{gi}^n})
    \end{split}\label{eq:rd_1}\\
    \begin{split} 
      & \boldsymbol{P_{gi}^n} -\Biggl(P_{gi}^o + K_i\,\sum_{i\in\mathcal{B}_{isl}}(\boldsymbol{P_{di}^n}-P_{di}^o) \\
      &  \qquad + K_i\left[\sum_{i\in\mathcal{G}_{isl}}(\boldsymbol{P_{gi}^n}-P_{gi}^o) + \Delta \right]\Biggr)\!\!\geq\!M(1\!-\!\boldsymbol{z_{gi}^n})
    \end{split}\label{eq:rd_2}\\
    & \sum_{i \in \mathcal{G}_{isl}} \boldsymbol{P_{gi}^n} = \sum_{i \in \mathcal{B}_{isl}} \boldsymbol{P_{di}^n}\\
    & \boldsymbol{z_{gi}^n}P_{gi}^{min} \leq \boldsymbol{P_{gi}^n} \leq \boldsymbol{z_{gi}^n}P_{gi}^{max} \label{eq:por_gl_2} \\
    & \text{for all buses in the island $i \in \mathcal{B}_{isl}$}\\
    & 0 \leq \boldsymbol{P_{di}^n} \leq P_{di}^o && \hspace{-10mm} \label{eq:ls_rd}
\end{align}
Here, the superscripts $o$ and $n$ refer to the generation and load before and after the redispatch respectively. Similar to the RAS-SCOPF, the generators are redispatched using a distributed slack bus model and participation factors $K_i$.

% In the case where all active generators in the island have $K_i = 0$, the problem \eqref{eq:obj_rd}- \eqref{eq:ls_rd} will be infeasible when the load in the island is less than the generation. In these situations, we redispatch the generation in the island as follows:
% \begin{enumerate}
%     \item Ramp each generator in the island up or down to match load and generation as closely as possible, constrained by the generation limits $P_g^{min}$ and $P_g^{max}$.
%     \item If, even after redispatching generators, the generation is more than the load, trip generators in the island, starting from the generators with the smallest capacity, until the generation is lesser than or equal to the load.
%     \item At this point, if the load is greater than the generation, shed load by multiplying all loads in the islands by a factor $\alpha = \sum_{i\in\mathcal{G}}P_{gi}/\sum_{i\in\mathcal{B}}P_{di}$
% \end{enumerate}

% When all active generators in the island have non-zero $K_i$,done using an optimization problem. Other wise follow the follwing steps
% \begin{enumerate}
% \item Ramp generators up/down to match demand in the island till you hit the limits of the generators.
% \item If generation is still more, trip generators until the load becomes greater than or equal to the generation.
% \item If load becomes greater, shed load by multiplying with factor.
% \end{enumerate}

%After redispatching the generators and loads, solve a DC power flow to compute the power flows in every transmission line in the island. 

\noindent \emph{\underline{Step 5: Trip overloaded lines}} We solve a DC power flow to compute the line flows and identify all lines in the island that are overloaded.
Because protective equipment for each line is assumed to have inverse time characteristics (i.e. it trips faster larger the overload), we trip only the line with the maximum overload, unless it is monitored by RAS and the RAS has not been triggered yet. Move to step 2). %the overloaded lines are not part of the set of lines monitored by the RAS. 

\section{Case Study} \label{sec:case-study}
The RAS-SCOPF is used to design a remedial action scheme for the 24-bus system shown in Fig. \ref{fig:rts96}. The system is based on the single area IEEE RTS-96 system described in \cite{rts96}, with some modifications. %When dispatched using the OPF, the original RTS-96 system is secure against all N-1 line outages. 
To make the case study more interesting, we reduce the line rating of all lines to 80\% of the original capacities listed in \cite{rts96}. This makes the system more congested, with a subset of contingencies causing post-contingency line overloads. %Since %, in the SCOPF and the RAS-aware SCOPF 
We only consider line outage constraints on non-radial lines and the capacity of the radial line 11 (from bus 7 to 8) is increased to 150\% of its original rating to ensure that it is never binding.  While designing the RAS, we consider only the peak load scenario with a total load of 2850 MW. We assume that only generators 1-16 are responsible for maintaining power balance in the system. For these generators, we define non-zero participation factors based on their maximum generation capacity,
\begin{equation}
    K_i = \frac{P_{gi}^{max}}{\sum_{k=1}^{16} P_{gk}^{max}} \quad\quad \text{ for } i=1,...,16.
\end{equation}
We assume a fixed penalty $\gamma$\,=\,\$5000/MW for load shedding and a fixed cost $\rho$\,=\,\$1000  for every generator that is shed from the network. We set $M=-m=100$ p.u.. %A smaller generator shedding cost was chosen because the priority of the RAS-SCOPF is to prevent load shedding when any contingency occurs in the system.
% \textcolor{red}{increased line limit at line 11}
% proposed in section \ref{sec:OPF}
% In this case study, we only consider N-1 line contingencies when designing and evaluating the RAS. For this case study, We are designing a RAS that monitors the flow on line 23 and protects the system from overloads in it. To identify which N-1 line outage results in an overload in line 23, we initially dispatch the system using the standard OPF. Then we subject the system to a contingency and recalculate the flows on the lines by solving a DC power flow. \textcolor{red}{Insert table of critical contingencies and percent overload}. 
% Table \ref{tab:cri_con} shows the contingencies that would result in overloads in line 23. This gives us the set $\mathcal{N}_c$ which corresponds to the outages of lines $\{7,18,21,22,27,29\}$.
%Table \ref{tab:cri_con} shows all the contingencies that would result in overloads in other lines in the system. 
% \begin{table*}[t]
%     \caption{Critical contingencies}
%     \begin{center}  
%     \begin{tabular}{|c|c|c|c|c|c|c|c|c|c|}
%     \hline
%         Outaged Line & 23 & 7 & 18 & 21 & 22 & 27 & 29 & 25 & 26 \\
%         % \hline
%         Overloaded Line & 7 & 23 & 23 & 23 & 23 & 23 & 23 & 28 & 28 \\
%         Percent Overload & 102.02 & 120.37 & 100.84 & 108.33 & 111.00 & 120.37 & 108.18 & 103.99 & 103.99 \\
%          \hline
%     \end{tabular}    
%     \end{center}
%     \label{tab:cri_con}
% \end{table*}

\subsection{RAS design}
\begin{figure}%[t]
    \centering
    \includegraphics[width = 0.9\linewidth]{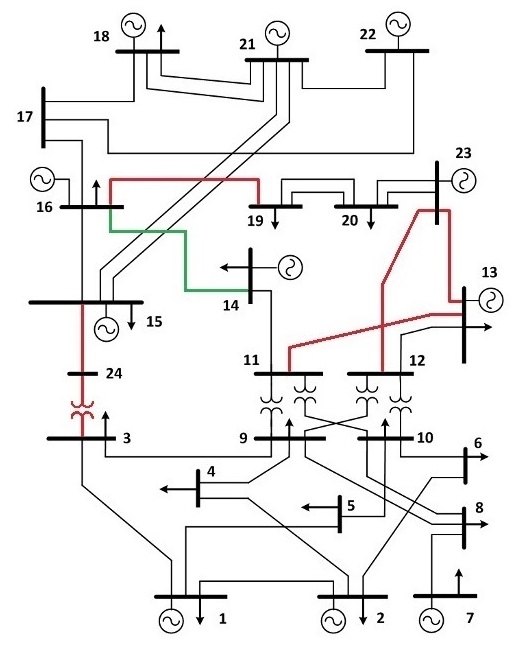}
    \caption{IEEE RTS-96 24-bus system. Line 23 is shown in green and the lines whose outages cause overloads in line 23 are shown in red}
    \label{fig:rts96}
\end{figure}

Table \ref{tab:cri_con} shows all the contingencies that would result in post-contingency overloads on other lines, assuming the initial dispatch is obtained using the OPF formulation.
% For the designed RAS, the set of monitored lines $\mathcal{L}_M = \{23\}$ and set of contingencies $\mathcal{N}_c = \{7,18,21,22,27,29\}$. 
% optimal threshold $\lambda_{RAS}^* = 0$ => if there is any overload trip the generators
% optimal set of generators to shed $\mathcal{G}_{shed} = \{17, 18, 19, 20, 22\}$
We observe that there are several different contingencies that lead to an overload on line 23. We
therefore demonstrate our proposed method by designing a RAS scheme that monitors overloads on this line. 
%For demonstration, we assume that the RAS to be designed only monitors line 23.
The set of contingencies $\mathcal{C}_M$ which the RAS is designed to protect against thus corresponds to the outages of lines $\{7,18,21,22,27,29\}$. Fig. \ref{fig:rts96} shows the location of the monitored line (ine green) and the lines that correspond to the critical contingencies (in red).

Solving the RAS-SCOPF with a single RAS scheme on line 23 and the given set of contingencies, we obtain a solution with pre-contingency generation cost \$62784.0. When detecting an overload in line 23, the RAS trips the generator $\mathcal{G}_{RAS} = \{22\}$, which corresponds to a 155 MW of generation capacity. The resulting power imbalance is balanced by generators 1-16, and does not cause any further overloads or load shed. As a result, we declare that our RAS scheme is effective. 
% The optimal value of the triggering threshold $\lambda_{RAS}$ is zero. This means that the RAS triggers whenever the flow on line 23 exceeds its rating.

%\textcolor{red}{make enumerate}
\subsection{Comparison with other OPF formulations}

We next compare the RAS-SCOPF to the OPF and SCOPF formulations described in Section II.B. We first solve each optimization problem and observe the corresponding pre-contingency generation costs and then evaluate the risk of post-contingency cascading failure by running cascading simulations for all contingencies listed in Table \ref{tab:cri_con}. 

Table \ref{tab:costs} shows the pre-contingency generation cost for the three OPF formulations. Dispatching generators using the OPF results in the lowest pre-contingency operating cost of the system. 
The RAS-SCOPF results in 3\% higher pre-contingency costs than the OPF, as it needs to ensure that the generator redispatch after the RAS action is feasible and that the initial dispatch is secure against critical contingencies that do not trigger the RAS. 
%The RAS-aware SCOPF procudes a solution that is approximately 3\% higher cost than the OPF, 
However, the cost of the RAS-SCOPF is significantly lower than the SCOPF, which results in a cost increase of nearly 12\%. This is because the SCOPF needs to ensure that all contingency constraints are satisfied without post-contingency generation tripping actions.  %results in a cost increase of nearly 12\%. %However, the reduction in operational cost comes at an increased risk of cascading failure. 
When considering the outcome of the cascading simulations for each solution, also listed in Table \ref{tab:costs}, we observe that the low-cost OPF solution has a significantly higher risk of cascading failure than the other two solutions. When the generators are dispatched using OPF, all the contingencies listed in Table \ref{tab:cri_con} result in cascading failure with a total load shed of 7832.8 MW across all contingencies. With the RAS-SCOPF and the SCOPF, none of the contingencies result in a cascading failure. %Thus, there is a trade-off between operational cost and security of the system. To ensure that the system is preventively secure, it must be operated at a higher cost.  
Based on these results, we conclude that the RAS we designed reduces operational costs relative to the SCOPF solution and lowers the risk of cascading failure relative to the OPF solution. %The pre-contingency costs obtained from the RAS-SCOPF is \$62784.0. Compared to the OPF, the RAS-SCOPF results in higher operation costs to ensure that (a) the generator redispatch after the RAS action is feasible and (b) the initial dispatch is secure against critical contingencies that do not trigger the RAS. Compared to the SCOPF, however, there is considerable reduction in operational costs. Additionally, the system is secure against all critical contingencies. Thus, RAS-SCOPF ensures system security at operational costs lower than that of the SCOPF.

% Cost is higher than that of OPF- because (a) fesible against contingencies which do not trigger the RAS (b) generator redispatch after RAS action must be feasible. However, when compared to SCOPF the cost is reduced. This benefit is while the system is secure against all contingencies

% This is slightly higher than the pre-contingency cost of the OPF, as the generators have to be dispatched to ensure that the generator re-dispatch after the RAS action is feasible. 

% When compared to the SCOPF, there is considerable reduction in operational costs. The system, however, is only secure against contingencies that were considered when designing the RAS. The occurrence of other critical contingencies results in cascading outages with a total load shed of 1290 MW. While the system is still not secure against all contingencies, the total load shed due to the critical contingencies can be reduced considerably by using an RAS. 

%  When the generators are dispatched using the RAS-aware SCOPF, the operational cost is \$63015.7, a 3.3\% increase relative to the OPF. Relative to the SCOPF, this solution represents a 7.6\% reduction, however the system still remains secure against all contingencies. Thus, RAS-aware SCOPF ensures system security at a reduced cost and thus can be used as an alternative to the SCOPF.
\begin{table}[t]
    \caption{Comparison of different OPF formulations}
    \begin{center}  
    \begin{tabular}{p{3.1cm}|C{1.2cm}C{1.2cm}C{1.2cm}}
        \hline
        & & & \\[-8pt]
        \textbf{Formulation} &  \textbf{OPF} & \textbf{RAS-SCOPF} & \textbf{SCOPF}\\
        \hline
        \hline
        & & & \\[-8pt]
         \textbf{Operational Cost} & \$61001.2 & \$62784.0 & \$68197.4\\
         \hline
        & & & \\[-7pt]
        \textbf{Cost increase relative to OPF (\%)} & 0\% & 2.92\% & 11.8\%\\
         \hline
         \hline
         & & & \\[-8pt]
         \textbf{Number of contingencies leading to load shed} & 9  & 0 & 0\\[+11pt]
         \hline
        & & & \\[-7pt]
         \textbf{Total load shed} & 7832.8 & 0 & 0\\
         \hline
    \end{tabular}    
    \vspace{-5mm}
    \label{tab:costs}
    \end{center} 
\end{table}

% \begin{table}[t]
%     \caption{Comparison of different OPF formulations}
%     \begin{center}  
%     \begin{tabular}{|C{2cm}|C{1.1cm}|C{1.1cm}|C{1.1cm}|C{1.1cm}|}
%         \hline
%         Formulation &  OPF & RAS-SCOPF & RAS-aware SCOPF & SCOPF\\
%         \hline
%          Cost & \$61001.2 & \$63103.1 &  \$63015.7 & \$68197.4\\
%          \hline
%          Percentage cost increase relative to OPF & 0\% & 1.15\% & 3.8\% & 11.8\%\\
%          \hline
%          Contingencies that lead to load shed & 9  & 2 & 0 & 0\\
%          \hline
%          Total Load Shed & 7832.8 & 1290  & 0 & 0\\
%          \hline
%     \end{tabular}    
%     \label{tab:costs}
%     \end{center} 
% \end{table}
%Modified RAS-SCOPF to include all critical contingencies, total cost = 63103.05. Secure against all contingencies. Seems to be slightly more conservative that the RAS-aware SCOPF for some reason

% Sensitivity to load- (0.9,1.1) case => few scenarios 1-4 range that are not secure out of 60-70
% higher deviations case average 17 failing scenarios and 13 overloads

\subsection{Sensitivity to load distribution}
 When designing the RAS, we only consider the peak load scenario. To assess how the RAS performs under different load conditions, cascade simulations are run for a range of loading conditions. To vary the load distribution, we first scale the load at every bus by a factor that is randomly drawn from a uniform distribution $X$, and then rescale the load to ensure that the total load in the system remains at 2850 MW,
 \begin{equation}  
    \begin{split}
        %& P_{di}^n = P_{di}^o X_i  \\
        & P_{di}^n = X_i P_{di}^o   \frac{\sum_{\ell \in \mathcal{B}}P_{d\ell}^o}{\sum_{j \in \mathcal{B}}P_{dj}^n}
      \end{split} \quad\quad \text{ for } i \in \mathcal{B}.
 \end{equation}
 
 When studying the performance of RAS when subjected to small load disturbances, $X \sim U(0.9,1.1)$, while $X \sim U(0.5,1.5)$ for larger load disturbances. %The load is then rescaled to ensure that the total load in the system remains at 2850 MW. 
 For each load scenario, the system is initialized by solving the RAS-aware SCOPF. Any scenario that renders the RAS-aware SCOPF infeasible is discarded. 
 
% To simulate smaller load disturbances, $X \sim U(0.9,1.1)$, while $X \sim U(0.5,1.5)$ for larger load deviations.

When load deviations are small ($X \sim U(0.9,1.1)$), around 70\% (68 out of 100) of scenarios have a feasible solution to the RAS-aware SCOPF. For most of the feasible load scenarios, the RAS is capable of preventing any critical contingency listed in table \ref{tab:cri_con} from evolving into a cascading event. However, there are 3 load scenarios where, for at least one critical contingency, the RAS action is not sufficient to remove the overload on line 23 resulting in a cascading failure. The total load shed across these scenarios is 1399.2 MW. 
% In all the feasible scenarios, the RAS is capable of preventing any initiating contingency from evolving into a cascading event. 

In the case of larger load deviation ($X \sim U(0.5,1.5)$), only 55\% (55 out of 100) of the scenarios have a feasible solution to the RAS-aware SCOPF. Among those scenarios, there are many that lead to cascading events after an initial line outage. Since outage of line 7 causes the largest overload in line 23, results pertaining only to the outage of line 7 are presented. The results are similar for all other contingencies in the set $\mathcal{C}_M$, except that fewer scenarios result in cascading failure. 

% Out of the 48 feasible scenarios,  6 scenarios result in cascading failure. 

Out of the 55 feasible scenarios, there are 16 scenarios where the RAS action is not sufficient to prevent cascading failure when line 7 is outaged. In all these cases, RAS was not able to remove the overload in line 23, resulting in overloads on other lines and eventually lead to cascading failure. For example, in one of the scenarios, the outage of line 7 triggers the RAS but the RAS action is insufficient to alleviate the overload on line 23. Thus, line 23 is tripped, followed by subsequent outages of lines 22 and 21 resulting in a total load shed of 1013.6 MW. The scenario described above was the worst case observed across all scenarios and contingencies.

% There are multiple ways in which the system can reach a state of cascade failure.  In 14 of the 15 cases of cascading failure, the RAS was not able to remove overload in line 23. For example, in one of these cases, the outage of line 7 triggered the RAS. However the actions of the RAS were not sufficient and line 23 remained overloaded. This resulted in line 23 being tripped, causing a cascading outages of line 21 and 22, in that order. When the simulation terminated, 844.3 MW of load had been shed and this was the worst case observed across all contingencies. In one case, the RAS action resulted in the overload of line 6. Here, the outage of line 7 triggered the RAS. The actions of the RAS, however, resulted in an overload in and the consequent outage of line 6. As a result, lines 2 and 11 are also tripped, resulting in a total load shed of 457.5 MW.

% There are 11 additional scenarios where lines other than line 23 get overloaded after the initiating contingency, but the system does not reach the state of cascading failure. In these cases, there can be load shed of as much as 14.5\%. While the system may not have satisfied our definition of system failure, it operates in a more vulnerable state with a substantial number of users disconnected from the supply.
% There are scenarios that lead to cascading outages, but the cascading terminates before the system reaches a state of failure (i.e, more than 10% buses get disconnected from the main network)
In 11 additional load scenarios, the outage of line 7 causes overloads in lines that are not monitored by the RAS. These overloads do not trigger the RAS, but cause cascading outages involving other lines. In these cases, % the cascading outages terminate before the system reaches a state of failure, i.e. more than 90\% of all buses remain connected to the largest island. Nevertheless, 
the system has separated into multiple islands and a significant amount load shed (up to 14.5\%) occurs. % and the system becomes more vulnerable to future contingencies. 

From these results, we could conclude that the RAS schemes either should be designed to be robust to large deviations in the loading condition, or that the RAS actions should be updated to reflect changing loading conditions using, e.g., the RAS-SCOPF.

\section{Conclusions}
\label{sec:conc}
Remedial action schemes (RAS) are an important tool to reduce congestion in power systems operation. However, RAS pose several challenges to grid operators due to their inability to adapt to changing conditions and the added risk of implementation failure. To address these challenges, we propose the RAS-SCOPF to choose a set of optimal RAS actions in response to current loading conditions. We further implement a DC power flow based cascading failure simulator to evaluate the risk of cascading failure when the RAS is present in the system and when it is not. 

The proposed method is applied to the RTS-96 24-bus network. Using the RAS-SCOPF lowered operational costs compared to the SCOPF, while ensuring same level of security against all critical contingencies. We further observe that the RAS designed using RAS-SCOPF is robust against small deviations in load, with very few scenarios resulting in system failure. However, the RAS was designed for only a single load scenario and is not effective in preventing cascading failures for loading scenarios that differ too much from the design scenario. This demonstrates the need for considering several load scenarios when solving the RAS-SCOPF or updating the RAS actions in real time. 

% In future work, we aim to address this by considering several load scenarios when solving the RAS-SCOPF. 
% However, because RAS-SCOPF is a mixed integer quadratic program (MIQP), it does not sale well as the system size and number of load scenarios and contingencies increase. Moreover, in our method, the RAS are considered to be fully reliable, i.e. their failure probabilities are assumed to be zero. In practice, however, the non-zero failure probabilities increase the operational risk and can lead to severe consequences. 
In future work, we aim to develop algorithms that solve the RAS-SCOPF efficiently for larger systems and multiple scenarios, while accounting for the RAS failure probabilities. 

\bibliographystyle{IEEEtran}
\bibliography{IEEEabrv, references}

\end{document}